\documentclass[12pt]{JHEP3}

\usepackage{amsfonts,amssymb}
\usepackage{epsfig}
\usepackage{varioref}
\title{MadEvent:
Automatic~event~generation~with~MadGraph\thanks{http://madgraph.physics.uiuc.edu}}

\author{Fabio Maltoni and Tim Stelzer\\
Department of Physics \\
University of Illinois at Urbana-Champaign \\ 1110 West Green Street \\
Urbana, IL\ \ 61801 \\
E-mail: \email{tstelzer@uiuc.edu} \\
E-mail: \email{maltoni@uiuc.edu} }

\abstract{We present a new multi-channel integration method and
its implementation in the multi-purpose event generator {\tt
MadEvent}, which is based on {\tt MadGraph}. Given a process, {\tt
MadGraph} automatically identifies all the relevant subprocesses,
generates both the amplitudes and the mappings needed for an
efficient integration over the phase space, and passes them to
{\tt MadEvent}. As a result, a process-specific, stand-alone code
is produced that allows the user to calculate cross sections and
produce unweighted events in a standard output format. Several
examples are given for processes that are relevant for physics
studies at present and forthcoming colliders.}

\preprint{hep-ph/0208156}

\begin{document}

\section{Introduction}
\label{sec:intro}

The high energy and luminosity of the present and future
colliders, from the Tevatron to the Large Hadron Collider to the
proposed 500 GeV Linear Collider, will offer the widest range of
physics opportunities for the exploration of the high-energy
frontier. Among the highest priorities is understanding not only
the nature of the electroweak symmetry breaking but also the
mechanism through which the electroweak scale stabilizes. Simple
and very robust arguments indicate that this scale should be less
than one TeV, very much in the reach of the above mentioned
colliders. At this energy, multijet final states in association
with leptons or missing energy constitutes the most interesting
data samples. The quest for the Higgs boson(s) and/or for
supersymmetric particles will rely on our ability to predict and
understand the standard-model processes which are the backgrounds
to these searches. In general, not only pure QCD interactions lead
to multijet configurations, but also the production of vector
bosons or top quarks that subsequently decay into quarks yield
similar signatures with many particles in the final state.

Although straightforward in principle, the evaluation of the cross
sections and especially the generation of unweighted events for
such processes is not a trivial task. First of all, the number of
subprocesses contributing to a process can be fairly large.  For
instance, for $W+4$-jet production in $pp$ collisions there are
$498$ distinct parton-parton subprocesses. Clearly, it is
desirable to automate not only the identification of these
subprocesses, but also their integration and event generation.

Unfortunately, the complexity of each subprocess also grows
quickly with the number of particles in the final state. If the
amplitudes involve identical particles then the growth is
factorial. A typical example, amply discussed in the
literature~\cite{Berends:1988me,Berends:1989ie,Mangano:1991by}, is
the computation of $n$-gluon amplitudes where already for a
relatively small number of gluons produced, say $n \gtrsim 6$,
clever reorganizations of the perturbative expansion have to be
employed to reduce the factorial growth to a more manageable
exponential one~\cite{Draggiotis:1998gr,Caravaglios:1998yr}.

The second step is the integration of the squared amplitude over the phase
space, which is necessarily performed using Monte Carlo techniques.
This task involves solving several issues. In the first place, the
amplitude results in a very complicated function of the momenta
which displays sharp peaks in many region of the phase space. In
order to achieve an efficient numerical integration, one needs to
identify the position of these peaks and to map them onto several
different sets of variables, which are usually referred to as
``channels''. In principle, both tasks are not trivial.
Fortunately, a standard Feynman diagram expansion provides the
necessary information on the structure (position and shape) of the
peaks in the matrix elements. In general, the main source of
variation of the amplitude in a given region of the phase space
can be associated with one (or more) of the propagators in one (or
more) diagrams becoming large. Typically, this happens either when
collinear-divergent limits are approached or when virtual massive
particles, with a small width, are close to their mass-shell.

At present, the available codes for the automated generation of
phase space are based on identifying singularities in Feynman
diagrams~\cite{Papadopoulos:2000tt, Krauss:2001iv}.
As we will show in the following, we use a
similar approach for generating the phase-space mappings.
However, our method differs from the previous ones in that
we propose to exploit the information gathered from the Feynman
diagrams to reorganize and simplify the
way we perform the integration over the various channels.

In this paper we present a new multi-channel integration
method, which we will call Single-Diagram-Enhanced, and
its implementation in the new multi-purpose event generator
{\tt MadEvent} which is powered by a
new version of {\tt MadGraph}~\cite{Stelzer:1994ta}.
Given a user process, {\tt MadGraph} automatically
generates the amplitudes for all
the relevant subprocesses and produces the mappings for the
integration over the phase space. This process-dependent
information is passed to {\tt MadEvent},
and a stand-alone code is produced that allows the user
to calculate cross sections and to obtain unweighted events.
Once the events have been generated (and
all the necessary information stored\footnote{
In addition to the usual kinematic quantities, such as masses and momenta,
the spin, the color connection, and the flavour of each of
the final-state partons may be needed for each event. This information
is stored in the ``Les Houches'' format~\cite{Boos:2001cv}.}) they may
be passed to any shower Monte Carlo program
(such as {\tt HERWIG}~\cite{Marchesini:1988cf} or
{\tt PYTHIA}~\cite{Bengtsson:1987kr,Sjostrand:1994yb})
where partons are perturbatevely evolved through the emission of
QCD radiation, and eventually turned into physical states (hadronization).

This paper  is organized as follows. In the next section we present
our method for mapping the phase space and integrating the squared amplitude.
In Section~\ref{sec:examples}, we provide some examples
of the results that can be obtained with {\tt MadEvent}. The
conclusions and a brief discussion of the outlook are addressed in
the last section.

\section{Single-Diagram-Enhanced multi-channel integration}
\label{sec:signal}

In this section we present our approach for integrating a generic
squared amplitude over the phase space of the final-state
particles. We first review the usual multi-channel integration method,
following closely Ref.~\cite{Kleiss:1994qy}, since this serves to establish
the notation and allows us to point out the issues relevant to the
subsequent discussion.

Let us consider $f(\vec{\Phi})$ the function to be integrated over
the set of phase-space variables $\vec{\Phi}$. Given $N$ the number
of particles in the final state, $\vec{\Phi}$ is of dimension
$d=3N-4$. In order to account for the complex structure of $f$
several mappings of the Monte Carlo variables $\vec{x} \in
[0,1]^d$ are introduced and each of them generates a different
probability density, $g_i(\vec{\Phi}), i=1,\ldots,n$, where $n$ is
the total number of channels.\footnote{Each probability density
$g_i(\vec{\Phi})$ is related to the mapping through $\vec{\Phi} =
G^{-1}_i(\vec{x})$, where $G_i(\vec{\Phi})$ is the primitive of
$g_i(\vec{\Phi})$.} The optimization of the relative importance of
the different channels is achieved by introducing a set of weights
$\alpha_i, \sum_{i=1}^n \alpha_i=1, \alpha_i \ge 0$, and defining
a total probability density $g(\vec{\Phi})=\sum_{i=1}^n \alpha_i
g_i(\vec{\Phi})$, which combines the weights of each channel and
is still correctly normalized. The weight of a phase space
point $\vec{\Phi}$ can be written as
$w(\vec{\Phi})=f(\vec{\Phi})/g(\vec{\Phi})$   and the result of
the Monte Carlo integration and its variance are:
\begin{eqnarray}
I        &=&\int d\vec{\Phi}\, g(\vec{\Phi})\, w(\vec{\Phi}) =
\int d\vec{\Phi} f(\vec{\Phi}) \,, \\
V(\alpha)&=&\int d\vec{\Phi}\, g(\vec{\Phi})\, w(\vec{\Phi})^2 = \int
d\vec{\Phi} \frac{f(\vec{\Phi})^2}{g(\vec{\Phi})}\,.
\end{eqnarray}
As made explicit in the notation, the result of the integration
does not depend on the choice of the $\alpha_i$ (which can even
differ iteration by iteration), while the variance does. The
aim of the multi-channel approach is to find the best set of $\alpha_i$
which minimizes the variance $V$. Several strategies can be
followed~\cite{Kleiss:1994qy}. Here, we just comment on a few important
aspects of this approach.

In the first place we note that for each point in the phase space,
independently from the actual channel chosen, the determination of
$g(\vec{\Phi})$ implies the computation of the contributions from
all channels.  As long as the number of channels is not very large
and/or the computational burden for calculating $g(\vec{\Phi})$ is
much smaller than that for $f(\vec{\Phi})$, this might not be a
serious problem. However, in general, this procedure is
inefficient. During the integration, the relative weight of each
channel changes due to the optimization of the
$\alpha_i$. In particular, even if some of the channels become
less and less important compared to others, the time spent to
calculate their contributions remains the same. Another point of
weakness is that, even if the evolution of each of the $\alpha_i$
can be chosen to be independent from the others, in practice  the
$\alpha_i$ are correlated. Predicting the behavior and/or
the stability of an integration with a large number of channels
is very difficult and jeopardizes the robustness of the
solution.\footnote{This aspect can in practice hamper the
possibility of using many channels if the mapping $g_i$ are
obtained by a stratified Monte Carlo method.}

We argue that there is a natural way to overcome the above
limitations and to calculate the integral of a squared amplitude
over the phase space in an efficient way. The idea is very
simple and is based on the following observation. Assume that the
function to be integrated could be written in terms of a basis of
$n$ functions $f_i$ such that:
\begin{eqnarray}
f=\sum_{i=1}^n f_i \qquad \;{\rm with }\; \qquad f_i\ge 0
\,,\qquad \forall \;\; i \,,
\label{eq:fbasis}
 \end{eqnarray}
and such that the peak structure of each $f_i$ can be efficiently
mapped by a single channel $g_i$. Then, the integration of $f$
reduces to:
\begin{eqnarray}
I &=& \int d\vec{\Phi} f(\vec{\Phi})
  = \sum_{i=1}^{n}  \int d\vec{\Phi}
  \,g_i(\vec{\Phi})\,\frac{f_i(\vec{\Phi})} {g_i(\vec{\Phi})}
  = \sum_{i=1}^{n} I_i \,,
\end{eqnarray}
{\it i.e.}, to a sum of $n$ independent integrations. For a
generic integration problem, the identification of such a basis
might be too difficult and the above method be simply not viable.
However, in the case at hand, a natural decomposition arises from
the physical content of the process. We propose to use the
following basis:
\begin{eqnarray}
f_i =  \frac{|A_i|^2}{\sum_i |A_i|^2}  |A_{\rm tot}|^2 \,,
\label{eq:SDEMCI}
 \end{eqnarray}
where $A_i$ is the amplitude corresponding to a single Feynman
diagram and $A_{\rm tot}=\sum_i A_i$ is the total amplitude. It is
clear that each $f_i$ in Eq.~(\ref{eq:SDEMCI}) satisfies the
conditions~(\ref{eq:fbasis}) and form a ``complete''
basis.\footnote{ Eq.~(\ref{eq:SDEMCI}) is analogous to the
procedure suggested by Odagiri~\cite{Odagiri:1998ep} to extract
the leading color behaviour of QCD amplitudes.} More importantly,
the peak structure of each $f_i$ is the same as of the
single squared amplitude $|A_i|^2$. Finding the suitable mapping
$g_i$ is therefore straightforward, since it can be derived from
the propagator structure of the corresponding Feynman diagram. We
will refer to the above algorithm as Single-Diagram-Enhanced
multi-channel integration.

Decomposing the integration of the amplitude into $n$ independent
integrations has immediate advantages. First, in contrast to
the standard multi-channel integration, the evaluation of the
weight in one channel does not require the computation of the others.
This entails that, from the statistics point of view, in the
Single-Diagram-Enhanced integration method
the complexity of the computation does not increase with
the number of channels.

Second, the Single-Diagram-Enhanced integration
is parallel in nature. Integrations in each
channel can be performed by using different resources and the
results combined only at the very end. Considering the latest
developments in this field, such as the availability of
PC farms at low cost, we reckon this as an alluring
property of this approach.

A third useful aspect of our method is that, given a target
accuracy of the integration, it is trivial to reweight the channels
so that those whose contributions to the total result is small
(large) are evaluated with a smaller (greater) number of Monte
Carlo points. One simply introduces the analogues of the
$\alpha_i$ as in a standard multi-channel method,
defining them as $\alpha_i= I_i/I$.
This realizes in practice the desirable property that the
computational time spent on a single channel should be
proportional to its relevance to the final result.

Another interesting property of the decomposition
(\ref{eq:SDEMCI}) is its intrinsic modularity. The number of the
diagrams to be included in the decomposition is completely
arbitrary and dictated only by convenience.  In general, the
number of channels will be less than the number of Feynman
diagrams. The first important example is when $k$ identical
particles are present in the final state and $n$ can be easily
reduced by a factor $k!$. Another case is when some diagrams do
not display any peaked behavior and are dropped from
Eq.~(\ref{eq:SDEMCI}). Alternately, some diagrams can be
efficiently grouped together if the form of the resulting $g_i$
is predicted by general arguments, such as radiation
coherence.

\section{MadEvent at work: some examples}
\label{sec:examples}

In this section, we present some results for processes relevant at
hadron colliders, which have  been obtained with {\tt MadEvent}.
The purpose is to show the versatility and the potential
of our approach and also to provide benchmark cross sections for
the user. Detailed information on how to use the
{\tt MadGraph/MadEvent} package together with a web interface
for generating the code for a given process is available at
{\tt http://madgraph.physics.uiuc.edu}.
In order to facilitate the comparison with other available codes
we have chosen a very simple set of cuts and conventions,
which are summarized in Table~\ref{tab:conventions}.
In particular, the cuts in transverse energy and
$\Delta R$ are left quite loose so that regions of the phase space
where the matrix elements peak are not left out and the
efficiency of our integration technique can be fully tested.~\footnote{In fact, at the LHC
such cuts are not sufficient to ensure a good behaviour of the
perturbative series for some of the cross sections shown, as it is apparent
by comparing the processes with $n$ jets with the corresponding ones with
$(n+1)$ jets.}

In Table~\ref{tab:one} we present the cross sections
involving the production one or more vector bosons in association
with jets, both at the Tevatron and LHC. Apart from the interest
they might have on their own, these processes contribute to some of the most
important backgrounds for top-quark studies ($W+$jets, $W b\bar b+$jets),
for the Higgs searches ($W^+W^-$+jets, $W b\bar b+$jets, $ZZ+$jets) and
for SUSY particles searches ($Z+$jets, with $Z\to\nu \bar \nu$). In the single
vector boson production, final states are leptons and so that the vector bosons are not
contrained to be on-shell. This entails, for instance, that cross sections
involving a $Z$ boson, correctly include the diagrams where
$\gamma^* \to \ell^+\ell^-$ and their interference with
$Z^* \to \ell^+\ell^-$.

For simplicity, in the case of double vector boson production, we present
the results where the vector bosons are produced on shell and are not decayed.
Nonetheless, we stress that {\tt MadEvent} correctly handles also the decays, where
non-trivial phase-space mappings have to be introduced.
An example is provided by $W^+W^-$ production  where contributions from
diagrams where $\gamma^*/Z \to W^+W^-\to \ell^+ \ell^- \nu \bar \nu$ appear,
which demand particular care. In this case, there are different
regions in the phase space which can give large contributions,
coming from the $Z$ and one of the $W$'s, or both $W$'s being on shell.
We have ensured that the procedure used for the mapping of the phase
space automatically accounts for all the possibilities,
so that no approximation is introduced.

When possible, we have compared our results with the ones obtained
by running available codes (such as {\tt
VECBOS}~\cite{Berends:1991ax}, {\tt Wbbgen}~\cite{Mangano:2001xp}
and {\tt MCFM}~\cite{Campbell:2002tg}) and found good agreement.
We plan to conduct a detailed set of comparisons with the recently
released {\tt Alpgen}~\cite{Mangano:2002ea} in the future.

Table~\ref{tab:two} contains the cross sections
for the production one or more heavy-quark pairs in association
with jets, both at the Tevatron and LHC. In the case of
$t\bar t b \bar b $ production we also give the cross sections
at order $\alpha_{\rm em}^2\alpha_{S}^2$.
These last ones constitute the irreducible backgrounds in the search
of the Higgs when it is produced in association with a top-quark pair.
Also, the cross section for two bottom-quark pairs in association
with jets are listed. In the case of two bottom-quark pairs we have
also performed a detailed comparison with the
results of Ref.~\cite{Tsuno:2002ce} and found good agreement.

In Table~\ref{tab:three} cross sections are given
for the production of an Higgs boson in association with a top-quark pair,
with jets (``$W$-boson fusion'') and in association with a $W$ boson,
both at the Tevatron and LHC.

\TABULAR[t]{|l|l|}{
\hline
                        & Tevatron: $p \bar p$ @ 2 TeV\\
\raisebox{1.5ex}[0cm][0cm]{colliders}   & LHC:      $p p     $ @ 14 TeV\\
\hline
 & $\alpha_S(M_Z)=0.1185$\\
{QCD and EW couplings and parameters}&
 $\alpha_{\rm em}(M_Z)=1/128.9$, $\sin^2 \theta_W=0.2312$\\
 & CKM matrix: diagonal \\
\hline
 p.d.f.   & CTEQ5L (parametric)\\
\hline
 scales   & $\mu_F=\mu_R=M_Z$ \\
\hline
 & $m_Z=91.188$ GeV, $\Gamma_Z=2.495$ GeV \\
 \raisebox{1.5ex}[0cm][0cm]{vector boson masses and widths}&$m_W=79.96$ GeV, $\Gamma_W=2.06$ GeV \\
\hline
 heavy-quark (pole) masses& $m_t=174.3$ GeV, $m_b=4.7$ GeV \\
\hline
 & $M_H=120$ GeV, $\Gamma_H=3.7\cdot 10^{-3}$ GeV\\
 \raisebox{1.5ex}[0cm][0cm]{Higgs parameters and couplings\footnotemark{} }& $y_t=0.97$, $y_b=0.017$
    \\
\hline
& $E^T>10$ GeV, $|\eta|<2.5$, $\Delta R>0.4$, \\
 \raisebox{1.5ex}[0cm][0cm]{kinematic cuts}& for all the final-state particles\\
 \hline
} {Couplings and conventions used in the calculation of the
benchmark cross sections. \label{tab:conventions} }

\TABULAR[p]{c|c|c|c|c|c}
{\hline\hline
 Process$+n$~jets            & $n$&     order  &  unit & Tevatron & LHC \\
 \hline
                      & 0&                       &            &   758    &  3850 (3450)        \\
                      & 1&                       &            &   182    &  1700 (1520)        \\
{$ e^{+} \nu_e \, (e^{-} \bar \nu_e)$ }
& 2& {$\alpha_{\rm em}^2 \alpha_S^n$}&{pb}    & 46.6 & 742 (642)\\
                      & 3&                         &      &   12.0    & 337 (279)         \\
                      & 4&                         &      &   3.19    & 156 (122)         \\
\hline
                      & 0&                               &       &   210       &  1000\\
                      & 1&                               &       &   46.2      &   398\\
{$  e^+ e^- $}
& 2& {$\alpha_{\rm em}^2 \alpha_S^n$}&{pb}                       &   12.6      &   179\\
                      & 3&                               &       &   3.30      &   79.0\\
                      & 4&                               &       &   0.871     &   35.1\\
\hline
                      & 0&                               &       &     427   & 2330 (1770)\\
{$ e^{+} \nu_e \, (e^{-} \bar \nu_e)\; b \bar b$}
& 1& {$\alpha_{\rm em}^2 \alpha_S^{n+2}$}& {fb}                  &     195   & 2950 (2330)\\
                      & 2&                               &       &     73.1  & 2600 (1980)\\
\hline
                      & 0&                               &       &    165    &  3880\\
{ $e^+ e^- b \bar b$ }
& 1& {$\alpha_{\rm em}^2 \alpha_S^{n+2}$}& {fb}                  &    79.3   &  3080\\
                      & 2&                               &       &    28.0   &  1770\\
\hline
                      & 0&                               &       &     9.28    & 46.3\\
{$ W^+ W^- $ }
& 1& {$\alpha_{\rm em}^2 \alpha_S^{n}$}& {pb}                    &     3.84    & 37.0\\
                      & 2&                               &       &     1.23    & 25.3\\
\hline
                      & 0&                               &       &     1.49    & 10.0 (7.25)\\
{$ W^+ (W^-) Z$}
& 1& {$\alpha_{\rm em}^2 \alpha_S^{n}$}& {pb}                    &     0.633   & 10.7 (7.31) \\
                      & 2&                               &       &     0.209   & 9.15 (6.40)\\
\hline
                      & 0&                               &       &     1.04    & 6.70\\
{$ Z Z$ }
& 1& {$\alpha_{\rm em}^2 \alpha_S^{n}$}& {pb}                    &     0.440   & 4.95\\
                      & 2&                               &       &     0.133   & 2.97\\

\hline\hline}
{Benchmark cross sections for single and double vector boson production in association with jets
at hadron colliders. The decays into leptons, $W \to e \nu_e$ and $\gamma^*/Z \to e^+ e^-$,
are included in the single vector boson cross sections
and cuts, as described in the text, are applied to all the final state particles (jets and leptons).
In the case of $WW,WZ,ZZ$ production, the vector bosons are not decayed and are produced on shell.
Statistical errors for all processes are at the percent level.
\label{tab:one}}

\TABULAR[t]{c|c|c|c|c|c}
{\hline\hline
 Process$+n$~jets            & $n$&          order     &  unit & Tevatron & LHC   \\
\hline
                      & 0&                               &       &   7.67   & 579   \\
                      & 1&                               &       &   3.53   & 762   \\
\raisebox{1.5ex}[0cm][0cm]{$ t \bar t $}& 2&
\raisebox{1.5ex}[0cm][0cm]{$\alpha_S^{n+2}$}
&\raisebox{1.5ex}[0cm][0cm] {pb}                                 &   1.24   & 660   \\
                      & 3&                               &       &   0.385  & 460   \\
\hline
                      & 0&                               &       &   832    & 15000 \\
                      & 1&                               &       &   115    &  3010 \\
\raisebox{1.5ex}[0cm][0cm]{$b \bar b $} & 2&
\raisebox{1.5ex}[0cm][0cm]{$\alpha_S^{n+2}$}
&\raisebox{1.5ex}[0cm][0cm]{nb}                                  &   29.0   &  1110\\
                      & 3&                               &       &   6.35   &   356\\
\hline
                                   & 0 &                 &       &   14.5   &  3890\\
& 1 & \raisebox{1.5ex}[0cm][0cm]{$\alpha_S^{n+4}$}       &
\raisebox{1.5ex}[0cm][0cm]{fb}                                   &   8.21   &  6440\\
\cline{2-6}
\raisebox{1.5ex}[0cm][0cm]{$t \bar t  b \bar b$ }
                                   & 0 &                 &       &   1.14   & 336\\
& 1 & \raisebox{1.5ex}[0cm][0cm]{$\alpha_{\rm em}^2 \alpha_S^{n+2}$} &
\raisebox{1.5ex}[0cm][0cm]{fb}                                   &   0.747  & 380\\
\hline
                                   & 0 &                    &    &    86.1      & 4050\\
& 1 & \raisebox{1.5ex}[0cm][0cm]{$\alpha_S^{n+4}$}&
\raisebox{1.5ex}[0cm][0cm]{pb}                                   &    41.0      & 546\\
\cline{2-6}
\raisebox{1.5ex}[0cm][0cm]{$b \bar b  b \bar b $ }
                                   & 0 &        &                &    676      & 16100\\

& 1 & \raisebox{1.5ex}[0cm][0cm]{$\alpha_{\rm em}^2 \alpha_S^{n+2}$} &
\raisebox{1.5ex}[0cm][0cm]{fb}                                   &   428       & 4090\\
\hline\hline}
{Benchmark cross sections for heavy-quark pair production in association with jets (light quarks and gluons)
at hadron colliders. The heavy quarks are produced on shell and are not
decayed. In $t \bar t b \bar b$ and $b \bar b b \bar b$ production, contributions from
the exchange of a virtual Higgs have been excluded.
Cuts, as described in the text, are applied to all the final state particles.
Statistical errors for all processes are at the percent level.
\label{tab:two}}


\TABULAR[b]{c|c|c|c|c|c}
{\hline\hline
 Process$+n$~jets             & $n$&          order                &  unit & Tevatron & LHC \\
\hline
                      & 2&                               &       &   157       &  1550\\
\raisebox{1.5ex}[0cm][0cm]{$h $} & 3&
\raisebox{1.5ex}[0cm][0cm]{$\alpha_{\rm em}^{3} \alpha_{S}^{n-2}$}
&\raisebox{1.5ex}[0cm][0cm]{fb}                                  &    89.3     & 1000 \\
\hline
                      & 0&                               &       &   7.30      &  545\\
{$ t \bar t  h $}& 1& {$ y_t^2 \alpha_S^{n+2}$}          & {fb}  &   3.14      &  830\\
                      & 2&                               &       &   1.00      &  852\\
\hline
                      & 0&                               &       &    67.9    &  563 (391)\\
                      & 1&                               &       &    29.3    &  425 (290)\\
\raisebox{1.5ex}[0cm][0cm]{$W^+ (W^-) h $} & 2& \raisebox{1.5ex}[0cm][0cm]{$\alpha_{\rm em}^{2} \alpha_{S}^{n}$}
&\raisebox{1.5ex}[0cm][0cm]{fb}                                  &     9.11   &  250 (168)\\
                      & 3&                               &       &     2.34   &  137 (89)\\
\hline\hline} {Benchmark cross sections for Higgs production in
association with a top-quark pair, with jets (EW processes) and in
association with a $W$ boson, at hadron colliders. The heavy
quarks and the Higgs and the $W$ are produced on shell and are not
decayed. Cuts, as described in the text, are applied to all the
final state particles. Statistical errors for all processes are at
the percent level. \label{tab:three}}


\section{Conclusions and Outlook}
\label{sec:conclusions}

In this paper we have introduced a new integration method, called
Single-Diagram-Enhanced multi-channel integration, and its implementation
in the multi-purpose event generator {\tt MadEvent}.

We have discussed how the problem of integrating over the phase space
amplitudes that involve a rather large number of particles in the
final state can be reduced to the integration of amplitudes
associated to single Feynman diagrams. The latter are much simpler
tasks and can be tackled by many of the techniques already present
in the literature.  Our method best applies to the
automated computation of cross sections (and, more importantly,
to the generation of unweighted events) for processes where many
different parton subprocess contribute, such as vector boson
production in association with jets at hadron colliders and many
others.

We have presented some applications of {\tt MadEvent},  a
multi-purpose event generator based on the above ideas, to studies
relevant at the present and forthcoming colliders.  All the
necessary steps to generate a set of unweighted events are
performed automatically. The user only inputs the  process and the
desired integrated luminosity or the total number of unweighted
events.

\footnotetext{In fact $\Gamma_H$ and $y_b$ do not
enter the results presented here, but we prefer to leave them as a reference for
the user.}

Although in principle there is no limitation to the number of
particles that can be considered in the final state, in practice
this still depends on the complexity of the process itself and on
the total number of Feynman diagrams involved. Currently, the
package is limited to ten thousands diagrams per subprocess. So,
for example, $W$+5 jets is feasible but close to the present
limit.

Several improvements that will extend the ability of {\tt MadGraph}
to deal with a larger number of QCD partons are under study.
One necessary step is the ``factorization'' of the amplitude
in order to reduce the complexity of the calculation from a
factorial growth to an exponential one, as already achieved for
non-QCD amplitudes in
Refs.~\cite{Caravaglios:1995cd,Ohl:2000hq,Kanaki:2000ey} and
extended to the QCD amplitudes in
Refs.~\cite{Draggiotis:1998gr,Caravaglios:1995cd, Mangano:2002ea}.
In the latter case, a reorganization of the color structure is also needed
and the sum over the colors of the external QCD partons can be efficiently
performed by a Monte Carlo integration. Work in this direction
is in progress~\cite{UIUCgang}.

\section*{Acknowledgments}

We thank Scott Willenbrock and Michelangelo Mangano for useful
discussions. We are also grateful to Henry Frisch, Bruce Knuteson
for their encouragement and Uli Baur and John Campbell for their
help in testing the package.

\bibliography{mad}
\end{document}